\documentclass[reprint,twocolumn,superscriptaddress,pre,showpacs,amsmath,amssymb,aps]{revtex4-1}

\usepackage{natbib}
\usepackage{graphicx,color,rotating}
\usepackage[latin1]{inputenc}
\usepackage{textcomp}
\usepackage{dcolumn}
\usepackage{bm}     

\newcommand{\notop}{{{}_{}}}

\renewcommand{\vec}[1]{\bm{#1}}
\newcommand{\ee}{\mathrm{e}}
\newcommand{\ii}{\mathrm{i}}
\newcommand{\dm}{\mathrm{d}}
\newcommand{\avr}[1]{\big\langle #1 \big\rangle}

\DeclareMathOperator{\re}{Re}

\newcommand{\iot}{{\ii\omega t}}

\newcommand{\pp}{\partial^{{}}}

\newcommand{\nablabf}{\boldsymbol{\nabla}}

\newcommand{\Lapl}{\nabla^2}
\newcommand{\nablasqr}{\nabla^2}

\newcommand{\ppt}{\partial^{{}}_t}
\newcommand{\ie}{\textit{i.e.}}

\newcommand{\etal}{\textit{et~al.}}

\newcommand{\scap}{\!\cdot\!}

\newcommand{\AperpMat}{\textbf{\textsf{A}}^{\perp_{}{}}}

\newcommand{\Aperpindex}{\textsf{A}^{\perp_{}{}}_{j,m}}
\newcommand{\eye}{\textbf{I}}
\newcommand{\aaa}{\vec{a}}

\newcommand{\ann}{a^\notop}

\newcommand{\bbb}{\vec{b}}

\newcommand{\bnn}{b^\notop}

\newcommand{\eee}{\vec{e}}
\newcommand{\een}{\vec{e}^\notop}

\newcommand{\eey}{\vec{e}^\notop_y}
\newcommand{\eez}{\vec{e}^\notop_z}

\newcommand{\FFF}{\vec{F}}

\newcommand{\FFFrad}{\vec{F}^\mathrm{rad}}
\newcommand{\Frad}{F^{\mathrm{rad}_{}}}

\newcommand{\nnn}{\vec{n}}

\newcommand{\vvv}{\vec{v}}

\newcommand{\vstra}{v^{{}}_\mathrm{str}}
\newcommand{\vstrT}{v^{T_{}}_\mathrm{str}}
\newcommand{\vvvstr}{\big\langle \vec{v}^{{}}_2 \big\rangle}

\newcommand{\vstry}{\big\langle v^{{}}_{2y} \big\rangle}
\newcommand{\vstrz}{\big\langle v^{{}}_{2z} \big\rangle}

\newcommand{\zerovec}{\boldsymbol{0}}

\newcommand{\aO}{a^{{}}_0}
\newcommand{\aOsqr}{a^{2_{}}_0}

\newcommand{\Apara}{A^{\parallel}}
\newcommand{\Aperp}{A^{\perp_{}}}

\newcommand{\cp}{c^{{}}_\mathrm{p}}

\newcommand{\Cp}{c^\notop_p}
\newcommand{\Cv}{c^\notop_v}

\newcommand{\Dth}{D^\notop_\mathrm{th}}

\newcommand{\Eac}{E^{{}}_\mathrm{ac}}

\newcommand{\FFFdrag}{\FFF^{\mathrm{drag}_{}}}

\newcommand{\kth}{k^\notop_\mathrm{th}}

\newcommand{\kapTi}{\tilde{\kappa}}

\newcommand{\KP}{\kappa^\notop_\mathrm{p}}
\newcommand{\kappas}{\kappa^\notop_s}

\newcommand{\Ta}{T^\notop_\mathrm{a}}

\newcommand{\upvec}{\vec{u}^\mathrm{p}}

\newcommand{\up}{u^\mathrm{p}}

\newcommand{\Dup}{\Delta \up}

\newcommand{\upexp}{\vec{u}^\mathrm{p_{}}_\mathrm{exp}}
\newcommand{\upanl}{\vec{u}^\mathrm{p_{}}_\mathrm{anl}}
\newcommand{\uradvec}{\vec{u}^\mathrm{rad}}

\newcommand{\urad}{u^\mathrm{rad}}

\newcommand{\Upp}{U^{{}}_\mathrm{pp}}

\newcommand{\va}{v^\notop_\mathrm{a}}
\newcommand{\vasqr}{v^{2_{}}_\mathrm{a}}
\newcommand{\vbc}{v^\notop_\mathrm{bc}}

\newcommand{\alphap}{{\alpha^\notop_p}}

\newcommand{\deltan}{\delta^\notop}

\newcommand{\deltaTi}{\tilde{\delta}}

\newcommand{\etaO}{\eta^{{}}_0}
\newcommand{\etaI}{\eta^{{}}_1}

\newcommand{\yti}{{\tilde{y}^{{}}{}}}
\newcommand{\zti}{{\tilde{z}^{{}}{}}}

\newcommand{\cO}{c^{{}}_0}

\newcommand{\fI}{f^{{}}_1}
\newcommand{\fII}{f^{{}}_2}

\newcommand{\pO}{p^{{}}_0}
\newcommand{\pa}{p^\notop_\mathrm{a}}

\newcommand{\pI}{p^{{}}_1}

\newcommand{\pII}{p^{{}}_2}

\newcommand{\TO}{T^{{}}_0}
\newcommand{\TI}{T^{{}}_1}

\newcommand{\uO}{u^{{}}_0}

\newcommand{\vvvI}{\vvv^{{}}_1}
\newcommand{\vII}{v^{{}}_2}
\newcommand{\vIIy}{v^{{}}_{2y}}
\newcommand{\vIIz}{v^{{}}_{2z}}
\newcommand{\vvvII}{\vvv^{{}}_2}

\newcommand{\rhoO}{\rho^\notop_0}
\newcommand{\rhoI}{\rho^\notop_1}

\newcommand{\rhoP}{\rho^\notop_\mathrm{p}}

\newcommand{\rhoTi}{\tilde{\rho}}

\newcommand{\SIC}{^\circ\!\textrm{C}}
\newcommand{\SICel}{^\circ\!\textrm{C}}

\newcommand{\SIMHz}{\textrm{MHz}}

\newcommand{\SIkg}{\textrm{kg}}

\newcommand{\SIm}{\textrm{m}}

\newcommand{\SImum}{\textrm{\textmu{}m}}
\newcommand{\SImu}{\textrm{\textmu{}}}

\newcommand{\SImus}{\textrm{\textmu{}s}}

\newcommand{\beq}[1]{\begin{equation} \eqlab{#1}}
\newcommand{\eeq}{\end{equation}}
\newcommand{\bsub}{\begin{subequations}}
\newcommand{\esub}{\end{subequations}}
\def\bal#1\eal{\begin{align}#1\end{align}}
\def\bsubal#1\esubal{\bsub \begin{align}#1\end{align} \esub}
\newcommand{\nn}{\nonumber}
\newcommand{\eqlab}[1]{\label{eq:#1}}
\renewcommand{\eqref}[1]{Eq.~(\ref{eq:#1})}
\newcommand{\eqsref}[2]{Eqs.~(\ref{eq:#1}) and~(\ref{eq:#2})}

\newcommand{\figref}[1]{Fig.~\ref{fig:#1}}

\newcommand{\figlab}[1]{\label{fig:#1}}

\newcommand{\secref}[1]{Section~\ref{sec:#1}}

\newcommand{\seclab}[1]{\label{sec:#1}}
\newcommand{\tabref}[1]{Table~\ref{tab:#1}}

\newcommand{\tablab}[1]{\label{tab:#1}}

\begin{document}

\title{Ultrasound-induced acoustophoretic motion of microparticles in three dimensions}%

\author{P. B. Muller}
\affiliation{Department of Physics, Technical University of Denmark, DTU Physics Building 309, DK-2800 Kongens Lyngby, Denmark}

\author{M. Rossi}
\affiliation{Universit\"{a}t der Bundeswehr M\"{u}nchen,
Werner-Heisenberg-Weg 39, 85579 Neubiberg, Germany}

\author{\'{A}. G. Mar\'{\i}n}
\affiliation{Universit\"{a}t der Bundeswehr M\"{u}nchen,
Werner-Heisenberg-Weg 39, 85579 Neubiberg, Germany}

\author{R. Barnkob}
\affiliation{Department of Physics, Technical University of Denmark, DTU Physics Building 309, DK-2800 Kongens Lyngby, Denmark}

\author{P. Augustsson}
\affiliation{Department of Measurement Technology and Industrial Electrical Engineering, Lund University, PO-Box 118, S-221 00 Lund, Sweden}

\author{T. Laurell}
\affiliation{Department of Measurement Technology and Industrial Electrical Engineering, Lund University, PO-Box 118, S-221 00 Lund, Sweden}
\affiliation{Department of Biomedical Engineering, Dongguk University, Seoul, South Korea}

\author{C. J. K\"{a}hler}
\affiliation{Universit\"{a}t der Bundeswehr M\"{u}nchen,
Werner-Heisenberg-Weg 39, 85579 Neubiberg, Germany}

\author{H. Bruus}
\affiliation{Department of Physics, Technical University of Denmark, DTU Physics Building 309, DK-2800 Kongens Lyngby, Denmark}
\email{bruus@fysik.dtu.dk}

\date{1 March 2013}

\begin{abstract}
We derive analytical expressions for the three-dimensional (3D) acoustophoretic motion of spherical microparticles in rectangular microchannels. The motion is generated by the acoustic radiation force and the acoustic streaming-induced drag force. In contrast to the classical theory of Rayleigh streaming in shallow, infinite, parallel-plate channels, our theory does include the effect of the microchannel side walls. The resulting predictions agree well with numerics and experimental measurements of the acoustophoretic motion of polystyrene spheres with nominal diameters of 0.537~$\SImum$ and 5.33~$\SImum$. The 3D particle motion was recorded using astigmatism particle tracking velocimetry under controlled thermal and acoustic conditions in a long, straight, rectangular microchannel actuated in one of its transverse standing ultrasound-wave resonance modes with one or two half-wavelengths.
The acoustic energy density is calibrated \textit{in situ} based on measurements of the radiation dominated motion of large 5-$\SImum$-diam particles, allowing for quantitative comparison between theoretical predictions and measurements of the streaming induced motion of small 0.5-$\SImum$-diam particles.
\end{abstract}



\maketitle


\section{Introduction}

Acoustofluidics is gaining increasing interest in lab-on-a-chip and microfluidics applications. Techniques based on acoustofluidic forces permit to perform a large variety of different tasks such as trapping, separation and sorting of cells, particle manipulation, and generation of fluid motion in a non-intrusive way~\cite{Bruus2011c, Laurell2007}. Acoustic forces allow for non-destructive and label-free particle handling based on size, density, and compressibility. Experimentally, the acoustophoretic motion of particles is driven by an ultrasonic standing wave that generates acoustic radiation forces on the particles and acoustic streaming in the fluid, which then exerts a Stokes drag force on the particles. Theoretically, such phenomena are described by complex, non-linear governing equations sensitive to the boundary conditions and are thereby difficult to predict. Therefore, the development of analytical and numerical methods that are able to accurately predict the acoustophoretic motion of different particle or cell types is currently a major challenge in the design of acoustofluidic systems \cite{Muller2012}.

To guide and control these theoretical developments, precise experimental measurements of the acousto\-pho\-retic motion of microparticles are necessary, and particle-based velocimetry techniques are among the best methods available. The work of Hag\-s\"{a}ter \etal\ \cite{Hagsater2007} was one of the first to use micro particle image velocimetry ($\SImu$PIV) in resonant microfluidic chips. In their case the measurements were employed to visualize the resonance modes in the microchip, using the radiation-dominated horizontal motion of 5-$\SImum$-diam particles and the associated horizontal acoustic streaming pattern using 1-$\SImum$-diam particles.
Using a similar $\SImu$PIV technique, Manneberg \etal\ \cite{Manneberg2009} characterized multiple localized ultrasonic manipulation functions
in a single microchip.
Barnkob \etal\ \cite{Barnkob2010} and Koklu \etal\ \cite{Koklu2010} also studied acoustophoretic motion of large particles (5- and 4-$\SImum$-diam, respectively), but instead used particle tracking velocimetry (PTV) to obtain particle paths, which were compared with theoretical results. Later, Augustsson \etal\  \cite{Augustsson2011} employed both PTV and $\SImu$PIV to make high-accuracy measurements of the acoustic energy density as well as the temperature and frequency dependence of acoustic resonances in microchannels filled with 5-$\SImum$-diam particles dominated by the radiation force. Such approaches have successfully been applied to the two-dimensional (2D) motion of particles in the optical focal plane in simple geometries and resonances. Recently, Dron \etal\ \cite{Dron2012} used defocusing of particle images to measure the magnitude of radiation-dominated acoustophoretic particle velocities parallel to the optical axis in similar simple half-wave resonance systems. However, in more complex configurations, or in the case of small particles dragged along by acoustic streaming rolls, more advanced techniques are necessary, that are able to resolve three-dimensional (3D) particle positions and three-component (3C) motion. Among these techniques, those based on $\SImu$PIV have issues regarding the depth of correlation between adjacent planes \cite{Olsen2000, Rossi2012}, while classical 3D particle tracking techniques require either stereo-microscopes with tedious calibration protocols, or fast confocal microscopes with a great loss in light intensity due to the use of pinholes \cite{Raffel2007}.

In this work, an analytical and experimental analysis is presented with the aim to improve the understanding of the full 3D character of ultrasound-induced acoustophoretic motion of microparticles. In particular, analytical expressions for this motion are obtained by extending the classical results for Rayleigh streaming in shallow parallel-plate channels to also cover rectangular channels of experimental relevance. The analytical results are compared with measurements of the 3D motion of particles in an acoustofluidic microchip performed by use of astigmatism particle tracking velocimetry (APTV)~\cite{Cierpka2010,Cierpka2011,Cierpka2012a}. APTV is a very precise single-camera tracking method which allows a time-resolved, volumetric reconstruction of the trajectories of microparticles in acoustophoretic motion. The technique is applicable to general 3D acoustophoretic motion of microparticles influenced by both the acoustic radiation force and the Stokes drag from acoustic streaming.

The paper is organized as follows. In \secref{theory} we derive an analytical expression of acoustic streaming in long, straight channels with rectangular cross-section, and we analyze the implications of this streaming for acoustophoretic motion of suspended microparticles. This is followed in \secref{experimental} by a description of the experimental methods, in particular the astigmatism particle tracking velocimetry technique. In \secref{results} we compare the theoretical and experimental results for the acoustophoretic microparticle motion, and finally in \secref{conclusions} we state our conclusions.

\section{Theory}
\seclab{theory}
The governing perturbation equations for the thermoacoustic fields are standard textbook material \cite{Morse1986, Pierce1991, Blackstock2000}. The full acoustic problem in a fluid, which before the presence of any acoustic wave is quiescent with constant temperature $\TO$, density $\rhoO$, and pressure $\pO$, is described by the four scalar fields temperature $T$, density $\rho$, pressure $p$, and entropy $s$ per mass unit as well as the velocity vector field $\vvv$. Changes in $\rho$ and $s$ are given by the two thermodynamic relations
 \bsub
 \eqlab{thermodyn}
 \bal
 \dm \rho &= \gamma\kappas\;\rho\:\dm p - \alphap\: \rho\:\dm T,\\
 \dm s &= \frac{\Cp}{T}\:\dm T - \frac{\alphap}{\rho}\:\dm  p,
 \eal
 \esub
which besides the specific heat capacity $\Cp$ at constant pressure also contain the specific heat capacity ratio $\gamma$, the isentropic compressibility $\kappas$, and the isobaric thermal expansion coefficient $\alphap$ given by
 \bsubal
 \gamma &= \frac{\Cp}{\Cv} = 1+\frac{\alpha_p^2\TO}{\rho^{{}}_0c^{{}}_p\kappas},\eqlab{materialparam_a}\\
 \kappas &= \frac{1}{\rho}\bigg(\frac{\pp\rho}{\pp p}\bigg)^{{}}_{\!\!s},\eqlab{materialparam_b}\\
 \alphap &= -\frac{1}{\rho}\bigg(\frac{\pp\rho}{\pp T}\bigg)^{{}}_{\!\!p}\eqlab{materialparam_c}.
 \esubal
The energy (heat), mass (continuity), and momentum (Navier--Stokes) equations take the form
 \bsub
 \eqlab{governing_equations}
 \bal
 \rho T\big[\pp_t +(\vvv\scap\nablabf)\big]s &=
 \sigma^{\prime}\!:\!\nablabf\vvv + \nablabf\scap(\kth\nablabf T),\\
 \ppt\rho &= -\nablabf\cdot(\rho \vvv),\\
 \rho\big[\ppt+\vvv\cdot\!\nablabf\big]\vvv &=  -\nablabf p + \nablabf\!\cdot\!\Big[\eta\big\{\nablabf\vvv+(\nablabf\vvv)^\mathrm{T}\big\}\Big]\!
 \nn\\
 & \qquad + (\beta-1)\nablabf(\eta\nablabf\scap\vvv),
 \eal
 \esub
where $\eta$ is the dynamic viscosity, $\beta$ is the viscosity ratio, which has the value 1/3 for simple liquids \cite{Morse1986}, $\kth$ is the thermal conductivity, and $\sigma^{\prime}$ is the viscous stress tensor.
As in Ref.~\cite{Muller2012}, we model the external ultrasound actuation through boundary conditions of amplitude $\vbc$ on the first-order velocity $\vvvI$ while keeping $T$ constant,
 \bsub
 \eqlab{BC_general}
 \bal
 \eqlab{BC_T1}
 T &= \TO, \text{ on all walls},\\
 \eqlab{BC_v1_noslip}
 \vvv &= \zerovec, \text{ on all walls},\\
 \eqlab{BC_v1_actu}
 \nnn\cdot\vvvI &= \vbc(y,z)\: \ee^{-\iot}, \text{ added to actuated walls}.
 \eal
 \esub
Here $\nnn$ is the outward pointing surface normal vector, and $\omega$ is the angular frequency characterizing the harmonic time dependence written using  complex notation.

\subsection{First-order fields in the bulk}
To first order in the amplitude $\vbc$ of the imposed ultrasound field we can substitute the first-order fields $\rhoI$ and $s^\notop_1$ in the governing equations \eqref{governing_equations} using \eqref{thermodyn}. The heat transfer equation for $\TI$, the kinematic continuity equation expressed in terms of $\pI$, and the dynamic Navier--Stokes equation for $\vvvI$, then become
 \bsub
 \eqlab{AcoustBasicEq}
 \bal
 \pp_t \TI &= \Dth\Lapl\TI + \frac{\alphap\TO}{\rhoO\Cp}\;\pp_t\pI,
 \\
 \pp_t\pI &= \frac{1}{\gamma\kappas}
 \Big[\alphap\pp_t\TI-\nablabf\scap\vvvI \Big],
 \\
 \pp_t \vvvI &=
 -\frac{1}{\rhoO}\nablabf \pI  + \nu\nabla^2 \vvvI +
 \beta\nu\:\nablabf(\nablabf\scap\vvvI).
 \eal
 \esub
Here, $\Dth = \kth/(\rhoO \cp)$ is the thermal diffusivity, and $\nu = \etaO/\rhoO$ is the kinematic viscosity. A further simplification can be obtained when assuming that all first-order fields have a harmonic time dependence $\ee^{-\iot}$ inherited from the imposed ultrasound field \eqref{BC_v1_actu}. Then $\pI$ can be eliminated by inserting Eq.~(\ref{eq:AcoustBasicEq}b), substituting $\pp_t\pI = -\ii\omega\pI$, into Eq.~(\ref{eq:AcoustBasicEq}a) and (c). Solutions of \eqref{AcoustBasicEq} describe the formation of thin thermoviscous boundary layers at rigid walls. In the viscous boundary layer of thickness
\bal
 \eqlab{delta}
\delta = \sqrt{\frac{2\nu}{\omega}},
 \eal
the velocity gradients are large, because the velocity field changes from its bulk value to zero at the walls across this layer \cite{Morse1986, Pierce1991, Blackstock2000}. In water at $\omega/(2\pi)=2~\SIMHz$ it becomes  $\delta \approx 0.4~\SImum$.

We focus on the transverse standing-wave resonance sketched in \figref{streaming}, which is established by tuning of $\omega$ in the time-harmonic boundary condition \eqref{BC_v1_actu} to achieve one of the resonance conditions $n\lambda^\notop_n/2 = w$, $n = 1, 2, 3,\ldots$, where $\lambda^\notop_n = 2\pi \cO/\omega^\notop_n$ is the acoustic wavelength of the $n$th horizontal resonance. The associated first-order fields $\vvvI$, $\pI$, and $\TI$ in the bulk of the channel take the form
 \bsub
 \eqlab{FirstOrderFields}
 \bal
 \eqlab{def_v1}
 \vvvI &= \va\sin(k^\notop_n y + n\pi/2)\:\ee^{-\mathrm{i} \omega_n t}\eey,\\
 \eqlab{def_p1}
 \pI &= \pa\cos(k^\notop_n y + n\pi/2)\:\ee^{-\mathrm{i} \omega_n t},\\
 \eqlab{def_T1}
 \TI &= \Ta\cos(k^\notop_n y + n\pi/2)\:\ee^{-\mathrm{i} \omega_n t},
 \eal
 \esub
where $k^\notop_n=2\pi/\lambda^\notop_n =n\pi/w$ is the wavenumber of the $n$th horizontal resonance, and the oscillation amplitudes of the first-order fields, indicated by subscript ``a'', are related through $\vert\va/\cO\vert \sim \vert \pa/\pO\vert \sim \vert\Ta/\TO\vert \ll 1$, with $\cO$ being the isentropic speed of sound in water. The spatial form of the standing-wave resonance is determined entirely by the resonance frequency and the geometry of the resonator, while its amplitude (here $\va \approx 10^4\: \vbc$ \cite{Muller2012}) is governed by the specific form of $\vbc$ and of the Q-factor of the resonance cavity. The acoustic energy density $\Eac$ is constant throughout the cavity and given by
\bal
\Eac = \frac{1}{4}\rhoO v_\mathrm{a}^2 = \frac{1}{4}\kappa^\notop_0 p_\mathrm{a}^2.
\eal

\begin{figure}[t]
\centering
\includegraphics[width=1\columnwidth]{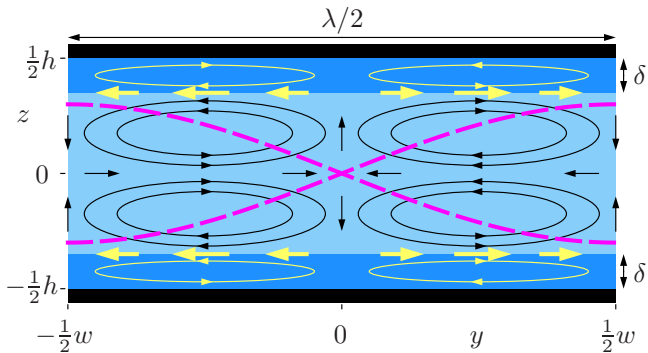}
\caption{\figlab{streaming} (Color online) A cross-sectional sketch in the $yz$-plane of the classical Rayleigh-Schlichting streaming pattern in the liquid-filled gap of height $h$ between two infinite, parallel rigid walls (black) in the $xy$-plane. The bulk liquid (light shade) supports a horizontal standing sinusoidal pressure half-wave $\pI$ (dashed lines) of wavelength $\lambda$ in the horizontal direction parallel to the walls. In the viscous boundary layers (dark shade) of sub-micrometer thickness $\delta$, large shear stresses appear, which generate the boundary-layer (Schlichting) streaming rolls (light thin lines). These result in an effective boundary condition $\avr{v^{\mathrm{bnd}_{}}_{2y}}$ (thick light arrows) with periodicity $\lambda/2$ driving the bulk (Rayleigh) streaming rolls (black thin lines). Only the top and bottom walls are subject to this effective slip boundary condition.}
\end{figure}

\subsection{Second-order governing equations for $\vvvstr$}
In a typical experiment on microparticle acoustophoresis, the microsecond timescale of the ultrasound oscillations is not resolved. It therefore suffices to treat only the time-averaged equations. The time average over a full oscillation period, denoted by the angled brackets $\avr{\cdots}$, of the second-order continuity equation and Navier--Stokes equation becomes
 \bsub
 \eqlab{SecondOrderNS}
 \bal
 \eqlab{SecondOrderCont}
 \rhoO\nablabf\cdot\vvvstr &= -\nablabf\cdot\avr{\rhoI\vvvI},\\[2mm]
 \etaO\Lapl\vvvstr &+\beta\etaO\nablabf(\nablabf\cdot\vvvstr)-\nablabf\avr{\pII}
 \nn\\
 &= \avr{\rhoI\pp_t\vvvI} + \rhoO\avr{(\vvvI\cdot\nablabf)\vvvI} \nn\\
 &\quad -\avr{\etaI\nablasqr\vvvI} - \avr{\beta\etaI\nablabf(\nablabf\cdot\vvvI)}
 \nn\\
 &\quad - \avr{\nablabf\etaI\cdot\left[\nablabf\vvvI + (\nablabf\vvvI)^\mathrm{T}\right]}
 \nn\\
 &\quad -\avr{(\beta-1)(\nablabf\cdot\vvvI)\nablabf\etaI}.
 \eal
 \esub
Here $\etaI$ is the perturbation of the dynamic viscosity due to temperature, $\eta = \etaO + \etaI = \eta(\TO) + \big[\pp_T\eta(\TO)\big]\:\TI$. From \eqref{SecondOrderNS} we notice that second-order temperature effects enter only through products of first-order fields. Dimensional analysis lead to a natural velocity scale $\uO$ for second-order phenomena given by
\bal
\uO = \frac{4\Eac}{\rhoO\cO} = \frac{\vasqr}{\cO}. \eqlab{uO}
\eal

\subsection{The boundary condition for bulk streaming flow}
The second-order problem \eqref{SecondOrderNS} was solved analytically by Lord Rayleigh \cite{LordRayleigh1884, Landau1993} in the isothermal case ($T = \TO$) for the infinite parallel-plate channel in the $yz$-plane with the imposed first-order bulk velocity $\vvvI$, \eqref{def_v1}. The resulting $y$-component $\avr{v^{\mathrm{bnd}_{}}_{2y}}$ of $\avr{\vvvII}$ just outside the boundary layers at the top and bottom walls becomes
 \bal
 \eqlab{v2bnd}
 \avr{v^{\mathrm{bnd}_{}}_{2y}} = -\vstra\: \sin\Bigg[n\pi\bigg(\frac{2y}{w}+1\bigg)\Bigg],
 \eal
as sketched in \figref{streaming} for the half-wave $k^\notop_1=\pi/w$. In Rayleigh's isothermal derivation the amplitude $\vstra$ of the streaming velocity boundary condition $\avr{v^{\mathrm{bnd}_{}}_{2y}}$  becomes
\bal
\eqlab{vstrO}
v^{0_{}}_\mathrm{str} = \frac{3}{8}\:\frac{\vasqr}{\cO} = \frac{3}{8}\uO,
\eal
where the superscript "0" refers to isothermal conditions. Recently, Rednikov and Sadhal \cite{Rednikov2011} extended this analysis by including the oscillating thermal field $\TI$ as well as the temperature dependence $\etaI(T)$ of the viscosity. They found that the amplitude of the streaming velocity boundary condition $\vstrT$ then becomes
 \bsub
 \eqlab{vstrT}
 \bal
 \vstrT &= \frac{8}{3} K^T v^0_\mathrm{str} = K^T \uO,\\
 K^T &= \frac{3}{8} + \frac{\gamma-1}{4}\Bigg[1-\frac{\big(\pp_T\eta\big)^{{}}_p}{\etaO\alphap}\Bigg] \frac{\sqrt{\nu/\Dth}}{1+\nu/\Dth},
 \eal
 \esub
where the superscript "$T$" refers to inclusion of thermoviscous effects leading to a temperature-dependent pre-factor multiplying the temperature-independent result. For water at $25~\SICel$ we find $\vstrT = 1.26\: v^{0_{}}_\mathrm{str}$ using the material parameter values of \tabref{parameters}, and in all calculations below we use this thermoviscous value for $\vstra$.

\subsection{Second-order governing equations for bulk $\vvvstr$}
In the bulk of the fluid the oscillating velocity and density fields $\vvvI$ and $\rhoI$ are out of phase by $\pi/2$. Consequently $\avr{\rhoI\vvvI}=0$, and the source term in the second-order continuity equation \eqref{SecondOrderCont} vanishes. As a result the time-averaged second-order velocity field $\vvvstr$ is divergence free or incompressible in the bulk. Hence, the continuity equation and the Navier--Stokes equation for the bulk streaming velocity field reduce to
 \bsub
 \eqlab{SecondOrderNSBulk}
 \bal
 \eqlab{avr_cont2}
 \nablabf\cdot\vvvstr &= 0,\\[2mm]
 \etaO\Lapl\vvvstr-\nablabf\avr{\pII}
 &= \avr{\rhoI\pp_t\vvvI} + \rhoO\avr{(\vvvI\cdot\nablabf)\vvvI}
 \nn\\
  &\quad - \avr{\etaI\nablasqr\vvvI} - \avr{\beta\etaI\nablabf(\nablabf\cdot\vvvI)}
 \nn\\
 &\quad -\avr{\nablabf\etaI\cdot\left[\nablabf\vvvI+(\nablabf\vvvI)^\mathrm{T}\right]}
 \nn\\
 \eqlab{avr_NS2}
 &\quad -\avr{(\beta-1)(\nablabf\cdot\vvvI)\nablabf\etaI}.
 \eal
 \esub
Only the $y$-component of the source terms on the right-hand side of \eqref{avr_NS2} is non-zero in the bulk, and it depends only on $y$ and not on $z$. Consequently, their rotation is zero, and they can be reformulated as a gradient term absorbed together with $\nablabf\avr{\pII}$ into an effective pressure gradient $\nablabf\chi$ given by,
 \bal
 \eqlab{effectivePressure}
 \nablabf\chi &=
 \nablabf\avr{\pII} + \avr{\rhoI\pp_t\vvvI} + \rhoO\avr{(\vvvI\cdot\nablabf)\vvvI}
  \nn\\
 &\quad - \avr{\etaI\nablasqr\vvvI} - \avr{\beta\etaI\nablabf(\nablabf\cdot\vvvI)}
 \nn\\
 &\quad - \avr{\nablabf\etaI\cdot\left[\nablabf\vvvI+(\nablabf\vvvI)^\mathrm{T}\right]}
 \nn\\
&\quad - \avr{(\beta-1)(\nablabf\cdot\vvvI)\nablabf\etaI}.
 \eal
Using this, the system of bulk equations reduces to the standard equation of incompressible creeping flow,
 \bsub
 \eqlab{SecondOrderNSBulkRed}
 \bal
 \nablabf\cdot\vvvstr &= 0,\\
 \eqlab{NSIIBulkRed}
 \etaO\Lapl\vvvstr &= \nablabf\chi.
 \eal
 \esub
These equations together with appropriate boundary conditions, to be discussed below, govern the steady bulk streaming velocity field $\avr{\vvvII}$ in the microchannel.

\subsection{Streaming in a parallel-plate channel}

Based on Rayleigh's analysis, we first study the analytical solution for $\avr{\vvvII}$ in the special case of a standing half wave ($n=1$) in the parallel-plate channel shown in \figref{streaming}. We choose the symmetric coordinate system such that $-w/2 < y < w/2$ and $-h/2 < z < h/2$, and introduce non-dimensionalized coordinates $\yti$ and $\zti$ by
 \bsub
 \begin{alignat}{2}
 \yti &= \frac{2y}{w}, && \text{ with } -1 < \yti < 1,\\
 \zti &= \frac{2z}{h}, && \text{ with } -1 < \zti < 1,\\
 \alpha &= \frac{h}{w},&&
 \text{ the aspect ratio}.
 \end{alignat}
 \esub
In this case, using \eqref{v2bnd}, the boundary conditions for $\avr{\vvvII(\yti,\zti)}$ are
 \bsub
 \eqlab{SecondOrderNSBulkBC}
 \begin{alignat}{2}
 \eqlab{vyTOP_sin}
 \vstry &= \vstra\: \sin(\pi\yti),
 &\; \mathrm{for}\ \zti &=\pm1,\\
 \eqlab{vzTOP_0}
 \vstrz &= 0,
 &\mathrm{for}\ \zti &=\pm1,\\
 \eqlab{vySIDE_0}
 \vstry &=0,
 &\mathrm{for}\ \yti &=\pm1,\\
 \eqlab{vzSIDE_sym}
 \partial^\notop_y \vstrz &= 0 ,
 &\mathrm{for}\ \yti &=\pm1,
 \end{alignat}
 \esub
where \eqsref{vySIDE_0}{vzSIDE_sym} express the symmetry condition at the wall-less vertical planes at $\yti = \pm1$.
\begin{table}[!t]
\caption{\tablab{parameters}Model parameters for water and polystyrene given at
temperature $T=25~\SIC$ and taken from the literature as indicated or derived based on these.}
\begin{ruledtabular}
\begin{tabular}{lccc}
Parameter & Symbol & Value & Unit \\ \hline
\hline
 \textbf{Water} & & & \\
Density\footnote{COMSOL Multiphysics Material Library \cite{comsol42a}.} & $\rho^{{}}_0$ & 998 & kg$\:$m$^{-3}$ \\
Speed of sound$^\mathrm{a}$ & $c^{{}}_0$ & 1495 & m$\:$s$^{-1}$ \\
Viscosity$^\mathrm{a}$ & $\eta$ & 0.893 & mPa$\:$s \\
Specific heat capacity$^\mathrm{a}$ & $c^{{}}_p$ & 4183 & J$\:$kg$^{-1}\:$K$^{-1}$ \\
Heat capacity ratio & $\gamma$ & 1.014 &  \\
Thermal conductivity$^\mathrm{a}$ & $k^{{}}_\mathrm{th}$ & 0.603 & W$\:$m$^{-1}\:$K$^{-1}$ \\
Thermal diffusivity & $D^{{}}_\mathrm{th}$ & 1.44 $\times$ 10$^{-7}$ & m$^2\:$s$^{-1}$ \\
Compressibility & $\kappa^{{}}_s$ & 448 & TPa$^{-1}$ \\
Thermal expansion coeff. & $\alpha^{{}}_p$ & 2.97 $\times$ 10$^{-4}$ & K$^{-1}$ \\
Thermal viscosity coeff.$^\mathrm{a}$ & $\frac{(\partial_T \eta)_p}{\eta^{{}}_0}$ & $-0.024$ & K$^{-1}$ \\
\hline
 \textbf{Polystyrene} & & & \\
Density\footnote{Ref.\ \cite{crc}.} & $\rho^{{}}_\mathrm{ps}$ & 1050 & kg$\:$m$^{-3}$ \\
Speed of sound\footnote{Ref.\ \cite{Bergmann1954}.} & $c^{{}}_\mathrm{ps}$ & 2350 & m$\:$s$^{-1}$ \\
Poisson's ratio\footnote{Ref.\ \cite{Mott2008}.} & $\sigma^{{}}_\mathrm{ps}$ & 0.35 &  \\
Compressibility\footnote{Calculated as $\kappa^{{}}_\mathrm{ps}=\frac{3(1-\sigma^{{}}_\mathrm{ps})}{1+\sigma^{{}}_\mathrm{ps}}\frac{1}{\rho^{{}}_\mathrm{ps} c_\mathrm{ps}^2}$, see Ref.~\cite{Landau1986}.} & $\kappa^{{}}_\mathrm{ps}$ & 249 & TPa$^{-1}$ \\
\end{tabular}
\end{ruledtabular}
\end{table}
Rayleigh focused his analysis of the parallel plate geometry on shallow channels for which $\alpha \ll 1$. Here $\alpha = 0.4$, derived from the aspect ratio of the microchannel described in \secref{experimental} and in Refs.~\cite{Barnkob2010, Augustsson2011, Barnkob2012a}, and consequently we need to solve the case of arbitrary $\alpha$. We find
 \bsub
 \eqlab{RayleighPlates}
 \bal
 \eqlab{v2y}
 \avr{\vIIy(\yti,\zti)} &= \vstra\sin(\pi\yti)\:\Apara(\alpha,\zti),\\
 \eqlab{v2z}
 \avr{\vIIz(\yti,\zti)} &= \vstra\cos(\pi\yti)\:\Aperp(\alpha,\zti),
 \eal
 \esub
where the $\alpha$- and $z$-dependent amplitude functions $\Apara$ and $\Aperp$ for the velocity component parallel and perpendicular to the first-order wave, respectively, are given by
 \bsub
 \eqlab{MullerPlatesz}
 \bal
 \Apara(\alpha,\zti) &=
 B(\alpha)\Big\{\big[1-\pi\alpha\coth(\pi\alpha)\big]\cosh(\pi\alpha\zti) \nn\\
 & \hspace*{30mm} + \pi\alpha\zti\sinh(\pi\alpha\zti)\Big\}, \eqlab{MullerPlateszPar}\\
 \Aperp(\alpha,\zti) &=
 \pi\alpha B(\alpha)\Big\{\coth(\pi\alpha)\sinh(\pi\alpha\zti) \nn\\
 & \hspace*{30mm} - \zti \cosh(\pi\alpha\zti)\Big\},\\
 B(\alpha) &= \frac{\sinh(\pi\alpha)}{\sinh(\pi\alpha)\cosh(\pi\alpha)-\pi\alpha},
 \eal
 \esub
with $\Apara(\alpha,\pm1) = 1$ and $\Aperp(\alpha,\pm1) = 0$.
In Rayleigh's well-cited shallow-channel limit $\alpha \ll 1$ the amplitude functions reduce to
 \bsub
 \eqlab{RayleighPlatesz}
 \begin{alignat}{2}
 \Apara(\alpha,\zti) &\approx
 \frac{3}{2}\zti^2-\frac{1}{2}, &
 \text{ for }  \alpha &\ll 1,\\[2mm]
 \Aperp(\alpha,\zti) &\approx \frac{\pi\alpha}{2}\big(\zti-\zti^3\big),&
 \text{ for }  \alpha &\ll 1.
 \end{alignat}
 \esub
The analytical solution of $\avr{\vvvII}$ for $\lambda/2 = w$ is illustrated in \figref{streaming_par_plates}(a) and (b)
for channel aspect ratios $\alpha = 1.2$ and 0.2. We note that the maximum streaming velocity is near the top and bottom walls. For the shallow channel \figref{streaming_par_plates}(b) there is furthermore a significant streaming velocity along the horizontal center line $\zti = 0$. However, the amplitude of this velocity decreases for increasing aspect ratio $\alpha$ as shown in \figref{streaming_par_plates}(d).

\begin{figure}
\centering
\includegraphics[width=0.99\columnwidth]{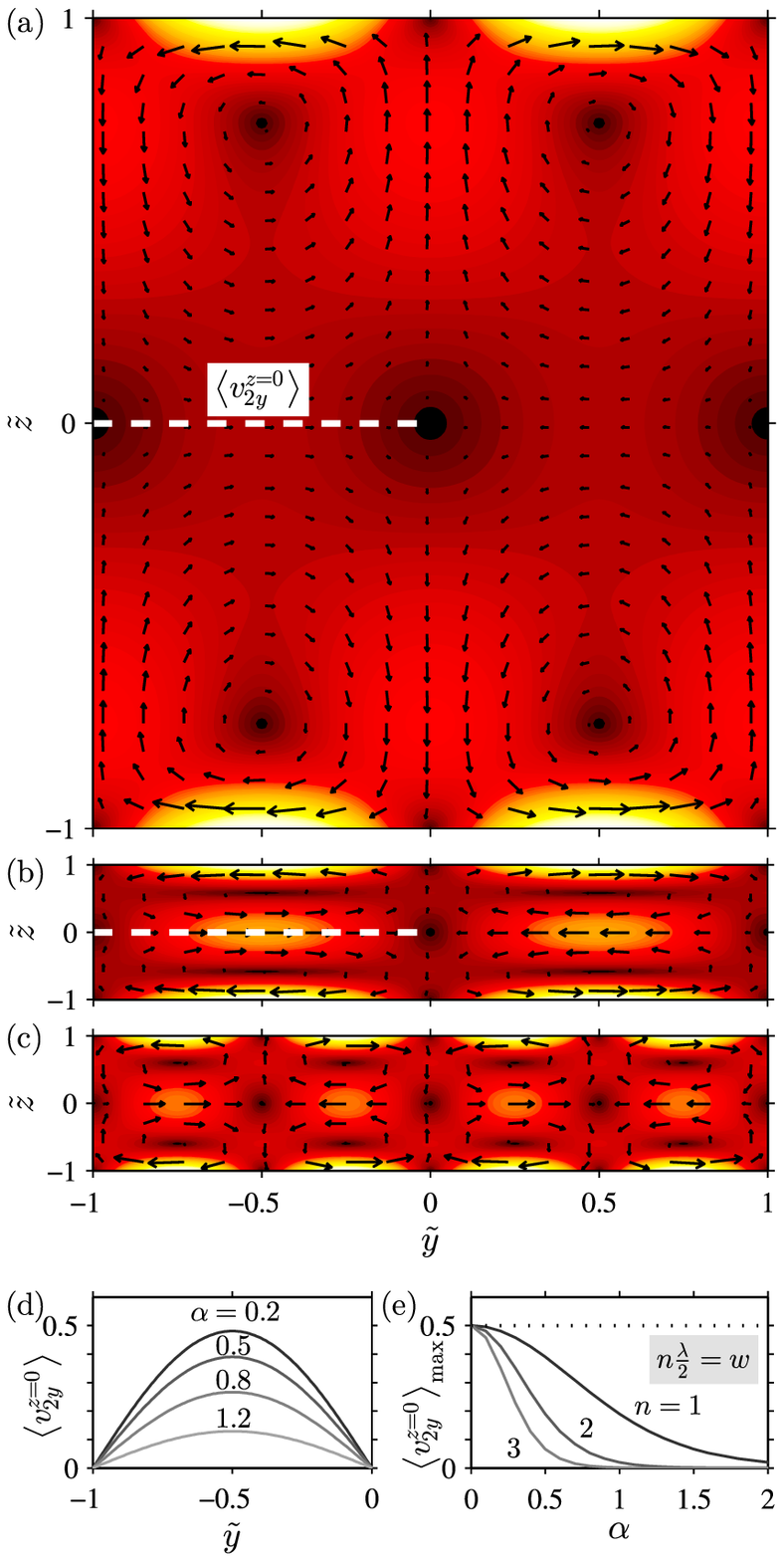}
\caption{\figlab{streaming_par_plates} (Color online) Analytical results for the streaming velocity $\avr{\vvvII}$ in parallel plate channels. (a) Plot of the analytical expressions  (\ref{eq:RayleighPlates}) and (\ref{eq:MullerPlatesz}) for $\avr{\vvvII}$ (arrows) and its magnitude (color plot from 0 (black) to $\vstra$ (white)) in the vertical $yz$ cross section of a parallel-plate channel (\figref{streaming}) with $\lambda/2 = w$ $(n=1)$ and aspect ratio $\alpha = 1.2$. (b) The same as (a), but for $\alpha = 0.2$. (c) The same as (b) but for a standing full wave, $\lambda = w$ $(n=2)$. (d) Line plot of the amplitude $\avr{v^{{}}_{2y}(\yti,0)}$ of the streaming velocity, in units of $\vstra$, along the first half of the center axis (white dashed lines in (a) and (b)) with $\lambda/2 = w$ for aspect ratios $\alpha = 0.2$, 0.5, 0.8, and 1.2. (e) Line plot of the maximum $\avr{v^{{}}_{2y}(\yti,0)}^{{}}_\mathrm{max}$ of the center-axis streaming velocity, in units of $\vstra$, as function of aspect ratio for the resonances $n\lambda/2 = w$, with $n = 1$, 2, and 3, respectively.}
\end{figure}%

This special case of the pure sinusoidal horizontal boundary condition \eqref{vyTOP_sin} can readily be generalized to any horizontal boundary condition by a Fourier expansion in wavenumber $k^\notop_m=2\pi/\lambda^\notop_m =m\pi/w$, where $m$ is a positive integer,
 \bsub
 \eqlab{vBC_Four}
 \bal
 \vstry &= \vstra\: f(\yti), \text{ for } \zti =\pm1, \\
 f(\yti) &= \sum_{m=1}^\infty \ann_m  \sin(m\pi\yti).
 \eal
 \esub
As the governing equations for the second-order bulk flow \eqref{SecondOrderNSBulkRed} are linear, we can make a straightforward generalization of \eqref{RayleighPlates}, and the two velocity components of the superposed solution for $\avr{\vvvII}$ become
 \bsub
 \eqlab{RayleighPlatesFour}
 \bal
 \avr{\vIIy(\yti,\zti)} &= \vstra\sum_{m=1}^\infty \ann_m \sin(m\pi\yti)\:\Apara(m\alpha,\zti),\\
 \avr{\vIIz(\yti,\zti)} &= \vstra\sum_{m=1}^\infty \ann_m \cos(m\pi\yti)\:\Aperp(m\alpha,\zti),
 \eal
 \esub
where the wave index $m$ multiplies both the horizontal coordinate $\yti$ and the aspect ratio $\alpha$. Note that $\Apara(m\alpha,\pm1) = 1$ and $\Aperp(m\alpha,\pm1) = 0$.
The resulting steady effective pressure $\chi$ is just the weighted sum of the partial pressures $\chi^{{}}_m$ of each Fourier component, $\chi = \sum_{m=1}^\infty \ann_m \chi^{{}}_m$.

In \figref{streaming_par_plates}(c) is shown the streaming velocity field for the higher harmonic boundary condition $f(\yti) = \sin(n\pi\yti)$ with $n=2$. Furthermore, \figref{streaming_par_plates}(e) shows how the maximum $\avr{v^{{}}_{2y}(\yti,0)}^{{}}_\mathrm{max}$ of the center-axis streaming velocity decays as function of aspect ratio $\alpha$ for $n = 1$, 2, and 3. Given sufficient room, the flow rolls decay in the vertical direction on the length scale of $\lambda^\notop_n/4$. Since $n$ is the number of half wavelengths of the first-order resonance pressure across the channel, we conclude that the streaming amplitude in the center of the channel decreases for higher harmonic resonances.

\subsection{Streaming in a rectangular channel}

Moving on to the rectangular channel cross section, we note that the only change in the problem formulation is to substitute the symmetry boundary conditions \eqsref{vySIDE_0}{vzSIDE_sym} by no-slip boundary conditions, while keeping the top-bottom slip boundary conditions \eqsref{vyTOP_sin}{vzTOP_0} unaltered,
 \bsub
 \eqlab{v2_BC_rect}
 \begin{alignat}{2}
 \eqlab{vyTOPrect_sin}
 \vstry &= \vstra\: \sin(\pi\yti),
 &\; \mathrm{for}\ \zti &=\pm1,\\
 \eqlab{vzTOPrect_0}
 \vstrz &= 0,
 &\mathrm{for}\ \zti &=\pm1,\\
 \eqlab{vySIDErect_0}
 \vstry &=0,
 &\mathrm{for}\ \yti &=\pm1,\\
 \eqlab{vzSIDErect_0}
 \vstrz &= 0 ,
 &\mathrm{for}\ \yti &=\pm1.
 \end{alignat}
 \esub

If we want to use the solution obtained for the parallel-plate channel, we need to cancel the vertical velocity component $\avr{\vIIz}$ on the vertical walls at $\yti=\pm1$. This leads us to consider the problem rotated $90^\circ$, where the first-order velocity field is parallel to the vertical walls (interchanging $y$ and $z$), and the fundamental wavelength is $\lambda/2 = h$, and the aspect ratio is $w/h = \alpha^{-1}$. As the governing equations for the bulk flow \eqref{SecondOrderNSBulkRed} are linear, we simply add this kind of solution to the former solution and determine the Fourier expansion coefficients such that the boundary conditions \eqref{v2_BC_rect} are fulfilled. Given this, \eqref{RayleighPlatesFour} generalizes to
 \bsub
 \eqlab{v2_rect_Four}
 \bal
 \eqlab{v2y_rect_Four}
 \avr{\vIIy(\yti,\zti)} = \vstra&\sum_{m=1}^\infty  \bigg[\ann_m \sin(m\pi\yti)\:\Apara(m\alpha,\zti)
 \nn\\
 & + \bnn_m \Aperp(m\alpha^{-1},\yti)\: \cos(m\pi\zti) \bigg],\\
 \eqlab{v2z_rect_Four}
 \avr{\vIIz(\yti,\zti)} = \vstra&\sum_{m=1}^\infty \bigg[\ann_m \cos(m\pi\yti)\:\Aperp(m\alpha,\zti)
 \nn\\
 & + \bnn_m \Apara(m\alpha^{-1},\yti)\:\sin(m\pi\zti) \bigg].
 \eal
 \esub

The two perpendicular-to-the-wall velocity conditions \eqsref{vzTOPrect_0}{vySIDErect_0} are automatically fulfilled as they by construction are inherited from the original conditions \eqsref{vzTOP_0}{vySIDE_0}. The unknown coefficients $\ann_m$ and $\bnn_m$ are thus to be determined by the parallel-to-the-wall conditions \eqsref{vyTOPrect_sin}{vzSIDErect_0}.

Using $\avr{\vIIy}$ in the form of \eqref{v2y_rect_Four}, boundary condition \eqref{vyTOPrect_sin} becomes
 \beq{v2y_rectTOP_BC}
 \sin(\pi\yti) =
 \sum_{m=1}^\infty  \bigg[\ann_m \sin(m\pi\yti)
  + (-1)^m \bnn_m \Aperp(m\alpha^{-1},\yti) \bigg].
 \eeq
The discrete Fourier transform of this equation, i.e. multiplying by $\sin(j\pi\yti)$, where $j$ is an arbitrary integer, and integrating over $\yti$ from $-1$ to $1$, becomes
 \beq{v2y_rectTOP_BC_Four}
 \deltan_{j,1}\! = \! \sum_{m=1}^\infty \!\Big[
  \deltan_{j,m}\:\ann_m \!+ \Aperpindex(\alpha^{-1})\: \bnn_m \Big],
 \; j = 1,2,3,\ldots
 \eeq
where the $(j,m)$'th element $\Aperpindex$ of the $\alpha$-dependent matrix $\AperpMat$ is given by
 \bal
 \eqlab{defFmatrix}
 \Aperpindex(\alpha^{-1})  &= (-1)^m\int_{-1}^{1} \!\dm \yti\:
 \Aperp(m\alpha^{-1},\yti) \: \sin(j\pi\yti).
 \eal
Introducing the coefficient vectors $\aaa$ and $\bbb$ and the first unit vector $\een_1$ with $m$'th components $\ann_m$, $\bnn_m$, and $\deltan_{1,m}$, respectively, we can write \eqref{v2y_rectTOP_BC_Four} as the matrix equation
\beq{v2y_rectTOP_BC_matrix}
\een_1 = \aaa + \AperpMat(\alpha^{-1})\cdot\bbb.
\eeq
Likewise, using \eqref{v2z_rect_Four} and multiplying it by $\sin(j\pi\zti)$, where $j$ is an arbitrary integer, and integrating over $\zti$ from $-1$ to $1$, the zero-parallel-component boundary condition \eqref{vzSIDErect_0} can be written as the matrix equation
\beq{v2y_rectSIDE_BC_matrix}
\zerovec = \AperpMat(\alpha)\cdot\aaa + \bbb.
\eeq
Solving the equation system \eqsref{v2y_rectTOP_BC_matrix}{v2y_rectSIDE_BC_matrix}, the coefficient vectors $\aaa$ and $\bbb$ becomes
 \bsub
 \eqlab{ab_coeff}
 \bal
 \eqlab{coeff_a}
 \aaa &= \bigg[ \eye - \AperpMat(\alpha^{-1})
  \AperpMat(\alpha)\bigg]^{-1}\!\!\cdot\een_1,\\
 \eqlab{coeff_b}
 \bbb &= -\AperpMat(\alpha)\cdot\aaa.
 \eal
 \esub
A comparison between results for the classical parallel-plate geometry and the new results for the rectangular geometry is shown in \figref{v2y_mid_Thigh_Trect}. It is seen how the velocity profile of the rectangular channel solution, \eqref{v2_rect_Four}, is suppressed close to the wall in comparison to the parallel-plate channel solution, \eqref{RayleighPlates}. Note that for the $n$th resonance, $k^\notop_n = n\pi/w$, the unit vector $\een_1$ in \eqref{coeff_a} is replaced by $(-1)^{n-1}\:\een_n$, with the sign originating from the $n$-dependent phase shift in the streaming boundary condition \eqref{v2bnd}.

\begin{figure}[t]
\centering
\includegraphics[width=1.0\columnwidth]{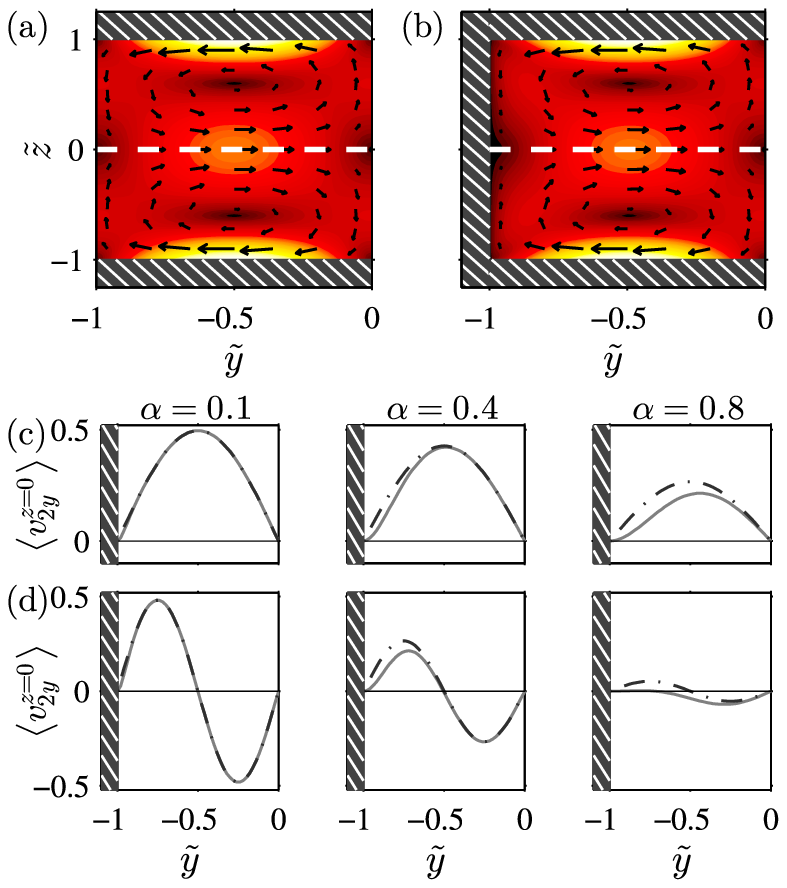}
\caption{\figlab{v2y_mid_Thigh_Trect} (Color online) Analytical results comparing the streaming velocity field $\avr{\vvvII}$ in the parallel plate and the rectangular channel. (a) Color plot from 0 (black) to $\vstra$ (white) of the analytical expression for $\avr{\vII}$ \eqsref{RayleighPlates}{MullerPlatesz} in the classical parallel-plate geometry with a half-wave resonance $\lambda/2 = w$ $(n=1)$. Due to symmetry, only the left half ($-1 < \yti < 0$) of the vertical channel cross section is shown. (b) As in (a) but for $\avr{\vII}$ in the rectangular channel \eqsref{v2_rect_Four}{ab_coeff}, including the first 20 terms of the Fourier series. (c) Line plots of $\avr{v_{2y}(\yti,0)}$ in units of $\vstra$ along the left half  of the center line for the parallel-plate channel (dashed lines) and the rectangular channel (full lines) for aspect ratios $\alpha = 0.1$, 0.4, and 0.8 and the half-wave resonance $\lambda/2 = w$. (d) As in (c) but for the full-wave resonance $\lambda = w$ $(n=2)$.
}
\end{figure}

\subsection{Acoustophoretic particle velocity}
\seclab{ParticleVelocity}
The forces of acoustic origin acting on a single microparticle of radius $a$, density $\rhoP$, and compressibility $\KP$ undergoing acoustophoresis with velocity $\upvec$ in a liquid of density $\rhoO$, compressibility $\kappas$, and viscosity $\etaO$, are the Stokes drag force $\FFFdrag = 6\pi\etaO a \big[\avr{\vvvII}-\upvec\big]$ from the acoustic streaming $\avr{\vvvII}$ and the acoustic radiation force $\FFFrad$. Given an observed maximum acoustophoretic velocity of $\up \lesssim 1$~mm/s for the largest particles of diameter $2a = 5.0~\SImum$, the Reynolds number for the flow around the particle becomes $\rhoO 2a \up/\eta \lesssim 6\times 10^{-3}$, and the time scale for acceleration of the particle becomes $\tau_\mathrm{acc} = \big[(4/3)\pi a^3 \rhoP \big]/\big[6\pi \eta a] \approx 2~\SImus$. Since the acceleration time is much smaller than the time scale for the translation of the particles $\tau_\mathrm{trans} = w/(2\up) \gtrsim 0.1$~s, the inertia of the particle can be neglected, and the quasi steady-state equation of motion, $\FFFdrag = -\FFFrad$, for a spherical particle of velocity $\upvec$ then becomes
 \beq{upvec}
 \upvec = \frac{\FFFrad}{6\pi\etaO a} + \avr{\vvvII} = \uradvec + \avr{\vvvII},
 \eeq
where $\uradvec$ is the contribution to the particle velocity from the acoustic radiation force.
The streaming velocity $\avr{\vvvII}$ is given in the previous subsections, while an analytical expression for the viscosity-dependent time-averaged radiation force $\FFFrad$ in the experimentally relevant limit of the wavelength $\lambda$ being much larger than both the particle radius $a$ and the boundary layer thickness $\delta $ was given recently by Settnes and Bruus \cite{Settnes2012}. For the case of a 1D transverse pressure resonance, \eqref{def_p1}, the viscosity-dependent acoustic radiation force on a particle reduces to the $x$- and $z$-independent expression
 \beq{Frad1D}
 \FFFrad(\yti) = 4\pi a^3 k^\notop_n\Eac\:\Phi(\kapTi,\rhoTi,\deltaTi)\:\sin\big[n\pi(\yti+1)\big]\eee^\notop_y.
 \eeq
The acoustic contrast factor $\Phi$ is given in terms of the material parameters as
 \bsubal
 \eqlab{PhiDef}
 \Phi(\kapTi,\rhoTi,\deltaTi) &=
 \frac{1}{3}\fI(\kapTi) + \frac{1}{2}\re\big[\fII(\rhoTi,\deltaTi)\big],\\
 \eqlab{f1RES}
 \fI(\kapTi) &=  1 - \kapTi,\\
 \eqlab{f2Res}
 \fII(\rhoTi,\deltaTi) &= \frac{2\big[1\!-\!\Gamma(\deltaTi)\big]
 (\rhoTi-1)}{2\rhoTi+1-3\Gamma(\deltaTi)},\\
 \eqlab{gamRES}
 \Gamma(\deltaTi) &=  -\frac{3}{2}
 \Big[1+\ii(1+\deltaTi)\Big] \deltaTi,
 \esubal
where $\kapTi = \KP/\kappas$, $\rhoTi = \rhoP/\rhoO$, and $\deltaTi = \delta/a$.
Using \eqref{Frad1D} for the transverse resonance, $\uradvec$ only has a horizontal component $\urad_y$
 \bsub
 \eqlab{urad_all}
 \beq{urady}
 \urad_y = \uO\: \frac{a^2}{\aOsqr}\:\sin\big[n\pi(\yti+1)\big], \; n = 1,2,3,\ldots,
 \eeq
where the characteristic particle radius $\aO$ is given by
 \beq{u0_a0}
  \aO = \delta\sqrt{\frac{3}{\Phi}},
 \eeq
 \esub
with $\delta$ given by \eqref{delta}.
The acoustophoretic particle velocity $\upvec$ will in general have a non-zero $z$-components, due to the contribution from the acoustic streaming $\avr{\vvvII}$. However, for the special case of particles in the horizontal center-plane $\zti =0$ of a parallel-plate or rectangular channel, the vertical streaming velocity component vanishes, $\avr{v_ {2z}(\yti,0)} = 0$. From \eqsref{v2y}{urady} we find that the horizontal particle velocity component $\up_y(\yti,0)$ in a parallel-plate channel is given by the sinusoidal expression,
  \beq{up}
  \up_y(\yti,0)  = \uO\bigg[\frac{a^2}{\aOsqr} - K^T\Apara(n\alpha,0)\bigg]
  \:\sin\big[n\pi(\yti+1)\big].
  \eeq
Since by \eqref{MullerPlateszPar} $\Apara(n\alpha,0)$ is always negative, it follows that the streaming-induced drag and the radiation force have the same direction in the horizontal center plane of the channel.
For the rectangular channel using \eqref{v2y_rect_Four}, the expression for $\up_y(\yti,0)$ becomes
  \bal
  \eqlab{upRect}
  &\up_y(\yti,0)  = \uO\Bigg\{\frac{a^2}{\aOsqr}\:\sin\big[n\pi(\yti+1)\big]
  \\ \nonumber &+
  K^T \sum_{m=1}^\infty  \bigg[\ann_m \sin(m\pi\yti)\:\Apara(m\alpha,0)
  + \bnn_m \Aperp(m\alpha^{-1},\yti) \bigg]\Bigg\},
  \eal
which is not sinusoidal in $\yti$ but still proportional to $\uO$. This particular motion in the ultrasound symmetry plane is studied in detail in Ref.~\cite{Barnkob2012a}.

\section{Experiments}
\seclab{experimental}
We have validated experimentally the analytical expressions derived above by measuring trajectories of micrometer-sized polystyrene particles displaced by acoustophoresis in a long, straight silicon/glass microchannel with rectangular cross section. A fully three-dimensional evaluation of the particle trajectories and velocities was performed by means of the astigmatism particle tracking velocimetry (APTV) technique~\cite{Cierpka2010,Cierpka2011} coupled to the
temperature-controlled and automated setup presented in Ref. \cite{Augustsson2011}. APTV is a single-camera particle-tracking method in which an astigmatic aberration is introduced in the optical system by means of a cylindrical lens placed in front of the camera sensor. Consequently,
a spherical particle image obtained in such a system shows a characteristic elliptical shape unequivocally related to its depth-position $z$. More details about calibration and uncertainty of this technique, as well as comparison with other whole-field velocimetry methods for microflows, can be found in Refs.~\cite{Cierpka2011, Cierpka2012a}.

\subsection{Acoustophoresis microchip}
The acoustophoresis microchip used for the experiment was the one previously presented in Refs.~\cite{Augustsson2011, Barnkob2010, Barnkob2012a}. Briefly, a rectangular cross section channel ($L = 35$~mm, $w = 377~\SImum$, and $h = 157~\SImum$) was etched in silicon. A Pyrex lid was anodically bonded to seal the channel and provided the optical access for the microscope. The outer dimensions of the chip are $L = 35$~mm, $W = 2.52$~mm, and $H = 1.48$~mm. From top and down, glued together, the chip was placed on top of a piezoceramic transducer (piezo), an aluminum slab to distribute heat evenly along the piezo, and a Peltier element to enable temperature control based on readings from a temperature sensor placed near the chip on the transducer. This chip-stack was mounted on a computer-controlled $xyz$-stage. Ultrasound vibrations propagating in the
microchip were generated in the piezo by applying an amplified sinusoidal voltage from a function generator, and the resulting piezo voltage $\Upp$ was monitored using an oscilloscope.

\subsection{APTV set-up and method}

The images of the particles in the microfluidic chip were taken using an epifluorescent microscope (DM2500 M, Leica Microsystems CMS GmbH, Wetzlar, Germany) in combination with a 12-bit, 1376$\times$1040 pixels, interline transfer CCD camera (Sensicam QE, PCO GmbH). The optical arrangement consisted of a principal objective lens with 20$\times$ magnification and 0.4 numerical aperture and a cylindrical lens with focal length $f_\mathrm{cyl} = 150$~mm placed in front of the CCD sensor of the camera. This configuration provided a measurement volume of $900\times 600\times 120~\SImum^3$ with an estimated uncertainty in the particle position determination of $\pm 1~\SImum$ in the $z$-direction and less than $\pm 0.1~\SImum$ in the $x$- and $y$-direction. Two scan positions along the $z$-direction were used to cover the whole cross-sectional area of the channel.

Monodisperse spherical polystyrene particles with nominal diameters of $5.33~\SImum$ (SD 0.09) and $0.537~\SImum$ (PDI 0.005) were used for the experiments ($\rho^{{}}_\mathrm{ps} = 1050~\SIkg~\SIm^{-3}$ and $\kappa^{{}}_\mathrm{ps} = 249~$TPa$^{-1}$). For simplicity we will refer to them as 5-$\SImum$-diam and 0.5-$\SImum$-diam particles, respectively. The particles were fabricated and labeled with a proprietary fluorescent dye by Microparticles GmbH to be visualized with an epifluorescent microscopy system. The illumination was provided by a continuous diode-pumped laser with 2~W at 532~nm wavelength (www.mylaserpage.de).

Once the particle 3D positions had been detected using the APTV technique, their trajectories and velocities were calculated. Due to the low seeding density in the experiments, the particle inter-distance was large enough to employ a simple nearest-neighbor approach in which the particle in one frame is identified with the closest particle in the next frame. The method was compared with more sophisticated ones as predictors and probabilistic algorithms with identical results. Trajectories composed by less than 5 particle positions were rejected.
From the obtained trajectories the velocities could be calculated given the capture rate of the camera. Different approaches have been followed depending on the type of trajectories expected. For particles following almost straight paths as those dominated by radiation force, a simple 2-position approach was used and the velocities were calculated based on the frame-to-frame particle displacement.
For particles with highly curved paths, like those present in streaming-dominated flows, a more sophisticated multi-frame approach has been followed, as those reported already by Hain and K\"{a}hler \cite{Hain2007} for $\SImu$PIV. In our case, each velocity data point was calculated from a trajectory segment composed by 4 consecutive points. Such multi-frame approach applied for PTV has been shown to better solve the velocity vector positions and values when the trajectories present large curvatures and for high-shear flows \cite{CierpkaLisbon}.

\section{RESULTS}
\seclab{results}

\subsection{APTV measurements}
\seclab{APTVmeas}

Examples of the measured 3D trajectories of the 5-$\SImum$-diam particles are shown in \figref{traject3D}(a). The data was collected from 10 consecutive experiments with the piezo operated at 1.94~MHz and peak-to-peak voltage of $\Upp = 0.91$~V. An overall number of 111 trajectories were determined. The 5-$\SImum$-diam particles are affected mainly by the acoustic radiation force $\Frad_y$ that quickly pushes them to the center of the channel with a horizontal velocity $\up_y$ \cite{Hagsater2007,Barnkob2012a}. At the vertical pressure nodal plane $y=0$, $\bm{F}^\notop_\mathrm{rad}$ vanishes and the hitherto negligible drag force from the acoustic streaming, shown in \figref{streaming_par_plates}(b), slowly drags the particles towards the top and bottom of the channel.

Examples of the measured 3D trajectories of the 0.5-$\SImum$-diam particles are shown in \figref{traject3D}(b). The data was collected from four consecutive experiments with the piezo operated at 1.94~MHz and peak-to-peak voltage of $\Upp = 1.62$~V. An overall number of 731 trajectories were determined. The acoustic radiation force $\Frad_y$ is in this case minute and the particles are primarily transported by the acoustic streaming $\vvvstr$ of the fluid resulting in particle trajectories following the four vertical vortices in the bulk, shown in \figref{streaming_par_plates}(b).

\begin{figure}
\begin{center}
 \includegraphics[width=1.0\columnwidth]{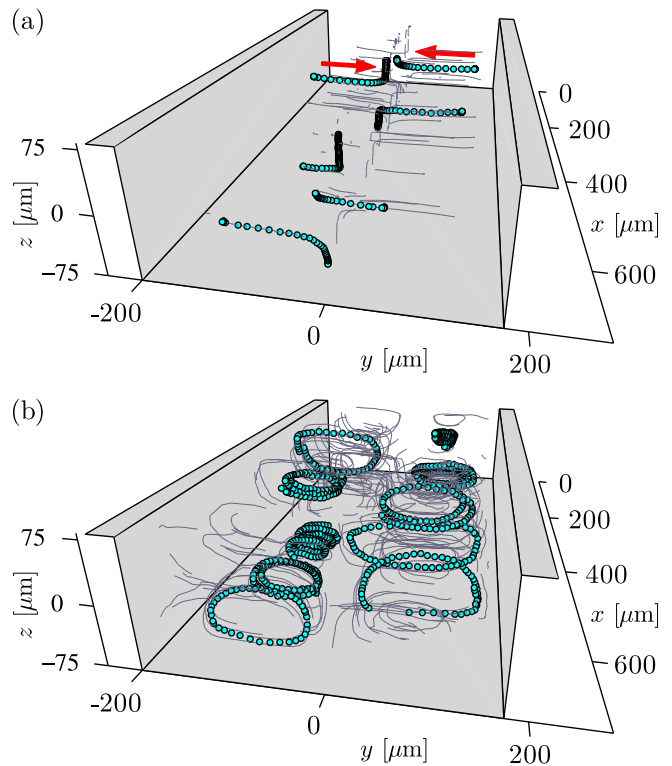}
\end{center}
\caption{\figlab{traject3D} (Color online) Measured particle trajectories (thin black lines) obtained using the 3D-APTV technique in the microchannel (gray walls) actuated at the 1.94-MHz horizontal half-wave resonance. For selected trajectories, the particle positions are represented by dots. (a) 5-$\SImum$-diam particles moving (red arrows) to the vertical center plane $y=0$, and (b) 0.5-$\SImum$-diam particles exhibiting circular motion as in \figref{streaming_par_plates}(b).}
\end{figure}

\subsection{Comparison of theory and experiments}
\begin{figure}[ht!]
\begin{center}
\includegraphics[width=1.0\columnwidth]{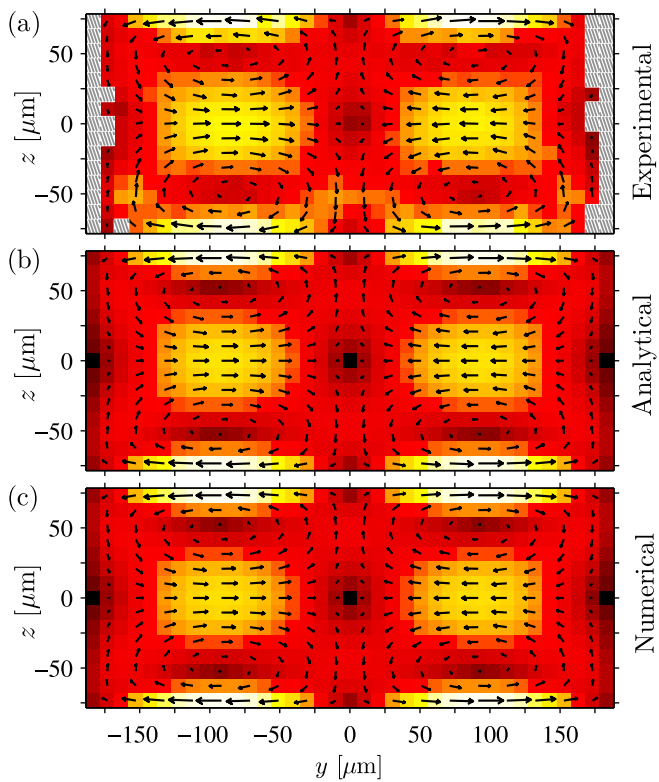}
\end{center}
\caption{\figlab{comp1} (Color online)
Comparison between experimental, analytical, and numerical studies of the acoustophoretic particle velocities $\upvec$ of 0.5-$\SImum$-diam polystyrene particles in water. The particle velocities $\upvec$ (vectors) and their magnitude (color plot ranging from 0 \SImum/s (black) to 63 \SImum/s (white) in all three plots), are shown in the vertical cross-section of the microchannel, divided into a pixel array consisting of 37-by-15 square bins of side length 10~\SImum. The axes of the plot coincide with the position of the channel walls. (a) The APTV measurements of the 0.5-$\SImum$-diam particles, shown in \figref{traject3D}(b), projected onto the vertical cross section. The maximum velocity is 63 \SImum/s. Close to the side walls experimental data could not be obtained, which is represented by hatched bins. (b) Analytical prediction of $\upvec$ based on \eqref{upvec}, taking both the radiation force and the streaming-induced drag force into account. The first 20 terms of the Fourier series for $\avr{\vvvII}$, \eqref{v2_rect_Four}, have been included in the calculation. The maximum velocity is 59 \SImum/s. There are no free parameters in this prediction as the acoustic energy density was calibrated \textit{in situ} based on measurements of large 5-$\SImum$-diam particles, shown \figref{traject3D}(a). (c) Numerical validation of the analytical result for $\upvec$ using the method described in Muller \etal\ \cite{Muller2012}. The numerical solution has been scaled by the thermoviscous pre-factor to the streaming amplitude \eqref{vstrT}. The maximum velocity is 59 \SImum/s.}
\end{figure}
Theoretically, the acoustophoretic particle velocity $\upvec$ is given by \eqref{upvec} combined with the expressions for the streaming velocity of the liquid \eqsref{v2_rect_Four}{ab_coeff} and the expression for the radiation force on the particles \eqref{Frad1D}.
The amplitudes of both the acoustic streaming and the radiation force depend linearly on the acoustic energy density $\Eac$ through \eqsref{vstrO}{Frad1D}.
To make a theoretical prediction of the motion of the 0.5-$\SImum$-diam particles we need to determine the acoustic energy density $E_\mathrm{ac}^{0.5~\SImum}$. This calibration is done \textit{in situ} based on the measurements of the 5-$\SImum$-diam particles, by the following three-step procedure.

First, we determine the acoustic energy density $E_\mathrm{ac}^{5~\SImum}$ for the experiment with the 5-$\SImum$-diam particles. This is done by fitting the $\sin(\pi\yti)$-dependent expression \eqref{up} for $\up_y(\yti,0)$ to the measured instantaneous velocities, using the amplitude as the only fitting parameter \cite{Barnkob2010, Barnkob2012a}. The small contribution from the acoustic streaming to the 5-$\SImum$-diam-particle velocity is taken into account although it constitutes only $6\:\%$ of the total particle velocity. The fit showed good agreement between theory and experiment, and after correcting for a wall-enhanced drag coefficient of 1.032 at the horizontal center plane (see Refs.~\cite{Koklu2010, Barnkob2012, Happel1983, Barnkob2012a}), we extracted the acoustic energy density $E_\mathrm{ac}^{5~\SImum} = (20.6\pm 0.7)~$J/m$^3$, where the $1\sigma$ standard error of the estimated value is stated. Since both the wall-enhanced drag coefficient and the drag force from the acoustic streaming fluid velocity are height-dependent, only five trajectories of 5-$\SImum$-diam particles close to the horizontal center line ($z=0$) qualified for use in the fit, based on a criterion of $|z^\notop_0|\leq6~\SImum$. The starting positions $(x^\notop_0,y^\notop_0,z^\notop_0)$ of the five tracks were $(34~\SImum,-115~\SImum,6~\SImum)$, $(310~\SImum,-66~\SImum,-6~\SImum)$, $(482~\SImum,-35~\SImum,-5~\SImum)$, $(74~\SImum,115~\SImum,2~\SImum)$, and $(350~\SImum,128~\SImum,0~\SImum)$, and they all reached the vertical center plane $y=0$.

Second, the acoustic energy density $E_\mathrm{ac}^{0.5~\SImum}$ for the experiment with the 0.5-$\SImum$-diam particles was determined, using the result for $E_\mathrm{ac}^{5~\SImum}$ combined with the fact that $\Eac$ scales as the square of the applied voltage $U^\notop_\mathrm{pp}$ \cite{Barnkob2010}.  The measured voltages for the two experiments are $U^{0.5~\SImum_{}}_\mathrm{pp} = (1.62 \pm 0.01)~V$ and $U^{5~\SImum_{}}_\mathrm{pp} = (0.91 \pm 0.01)~V$, where the stated error corresponds to the standard deviation of a series of voltage measurements, with the power turned off in between each measurement. The derived value for $E_\mathrm{ac}^{0.5~\SImum}$, taking into account the errors of $U^{0.5~\SImum_{}}_\mathrm{pp}$, $U^{5~\SImum_{}}_\mathrm{pp}$, and $E_\mathrm{ac}^{5~\SImum}$, becomes $E_\mathrm{ac}^{0.5~\SImum} = (U^{0.5~\SImum_{}}_\mathrm{pp}/U^{5~\SImum_{}}_\mathrm{pp})^2 E_\mathrm{ac}^{5~\SImum} = (65\pm 2)~$J/m$^3$, with the contribution from the error of the measured voltages being negligible.

Third, based on \eqref{uO}, the derived value for the energy density $E_\mathrm{ac}^{0.5~\SImum}$ is used in the analytical expression for the particle velocities \eqref{upvec}. The radiation force is given by \eqref{Frad1D} and the acoustic streaming velocity is given by \eqsref{v2_rect_Four}{ab_coeff}, using the thermoviscous-corrected amplitude $\vstrT$ \eqref{vstrT}. The contribution from the acoustic radiation force to the 0.5-$\SImum$-diam-particle velocity is small and constitutes only $12\:\%$ of the total particle velocity in the horizontal center plane $z=0$. The contribution from the radiation force to the 0.5-$\SImum$-diam-particle velocity is not corrected for the wall-enhanced drag coefficient, since this is minute for these small particles.

\begin{figure}
\begin{center}
\includegraphics[width=\columnwidth]{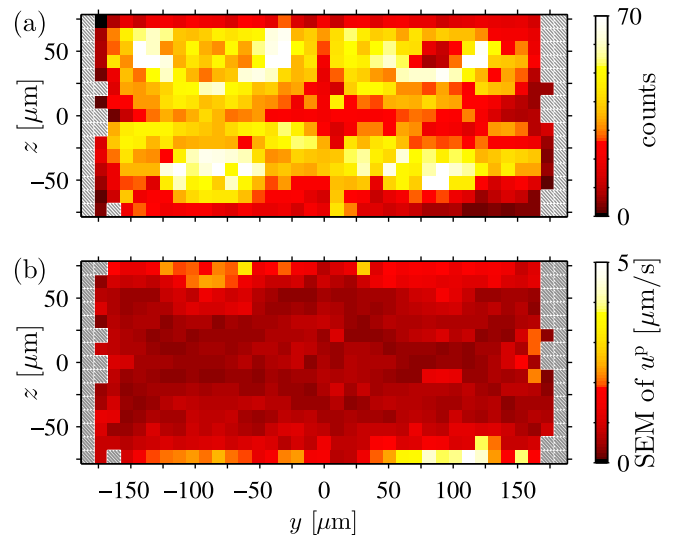}
\end{center}
\caption{\figlab{MeasStats} (Color online) (a) Color plot of the number of times the velocity has been measured in each square bin. (b) Color plot of standard error of the mean (SEM) particle velocity in each square bin.}
\end{figure}

To compare the experimental results and the analytical prediction, we consider the 0.5-$\SImum$-diam-particle velocities in the vertical cross section, $yz$-plane, of the channel as in Fig. \ref{fig:streaming}, \ref{fig:streaming_par_plates}, and \ref{fig:v2y_mid_Thigh_Trect}. In \figref{comp1} are shown color plots of (a) the experimentally measured acoustophoretic velocities for the 0.5-$\SImum$-diam particles, (b) the analytical prediction of the same, and (c) the numerical validation of the analytical result using the methods of Muller \etal\ \cite{Muller2012}. The three data sets are shown on the same $37\times 15$ bin array and with the same color scale. The experimental and the analytical velocities agree well both qualitatively and quantitatively, although the experimental velocities are approximately $20\:\%$ higher on average.
The experimental results for the particle velocities, \figref{comp1}(a), is found as the mean of several measurements of the particle velocity in each bin. The number of measurements performed in each bin is shown in \figref{MeasStats}(a), while the standard error of the mean (SEM) particle velocity is shown in \figref{MeasStats}(b). These plots show that we typically have between 20 and 70 measurements in each bin and the experimental error is on average only $1~\SImum/s$, while the relative experimental error is on average $4\:\%$. The error of the theoretical prediction is given by the relative error of $4\:\%$ on the estimated value for the energy density $E_\mathrm{ac}^{0.5~\SImum}$.
\begin{figure}
\begin{center}
\includegraphics[width=\columnwidth]{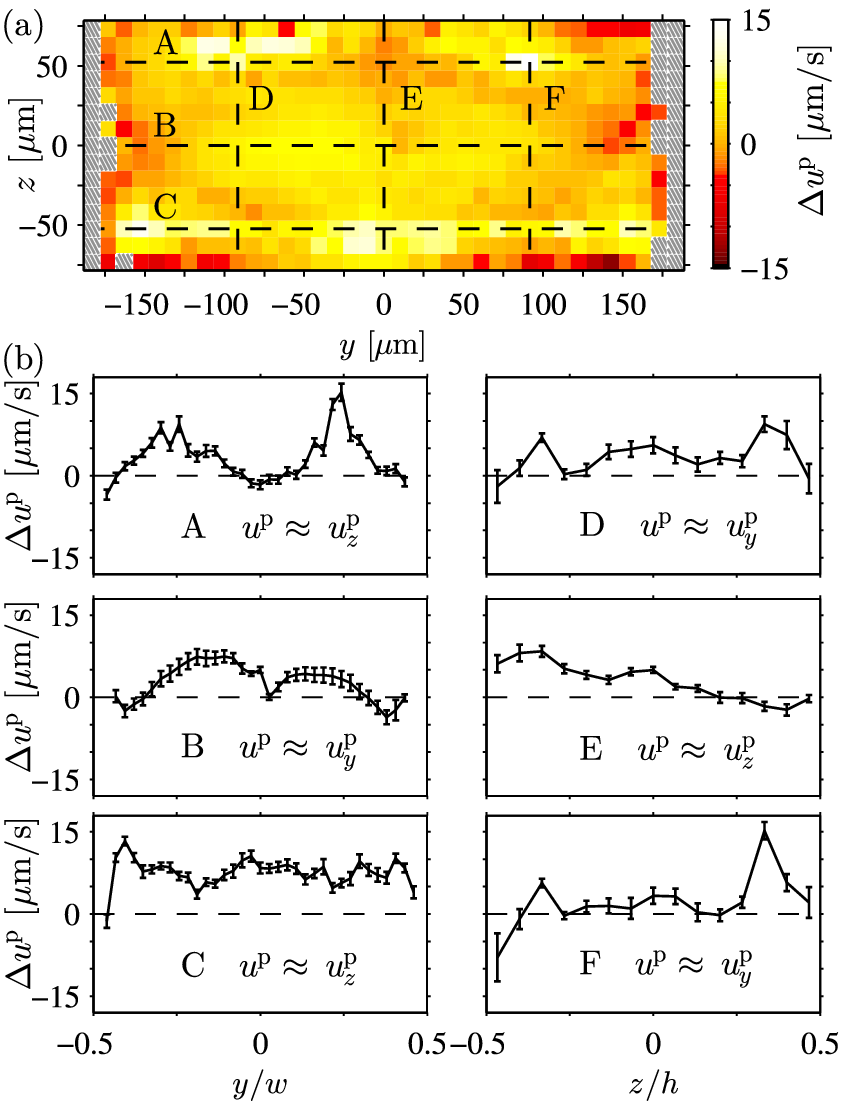}
\end{center}
\caption{\figlab{comp2} (Color online) (a) Color plot of the difference between the experimental and analytical acoustophoretic particle speeds, $\Dup$ \eqref{Dup}. (b) Line plots of $\Dup$ along the dashed lines in (a), marked A, B, C, D, E, and F, with error bars indicating the $1\sigma$ error of $\Dup$. The lines are positioned at $y=0~\SImum$, $y=\pm91.7~\SImum$, $z=0~\SImum$, and $z=\pm52.3~\SImum$. The off-center lines go through the rotation centers of the flow rolls, and consequently $\upvec\approx \up_y\eey$ in B, D, and F, while $\upvec\approx \up_z\eez$ in A, C, and E.}
\end{figure}

The quantitative differences between the experimental particle velocities \figref{comp1}(a) and the analytical prediction \figref{comp1}(b) are emphasized in \figref{comp2}, showing the difference $\Dup$ between the experimental and analytical acoustophoretic particle speeds
\bal
\eqlab{Dup}
\Dup = |\upexp| - |\upanl|.
\eal
We have chosen to consider the difference of the absolute velocity values, $|\upexp| - |\upanl|$, instead of the absolute of the difference, $|\upexp - \upanl|$, because the former allows us to see when the experimental velocity respectively overshoots and undershoots the analytical prediction.
\figref{comp2}(a) shows a color plot of $\Dup$ in the channel cross section, while \figref{comp2}(b) shows line plots of $\Dup$ along the dashed lines in \figref{comp2}(a), allowing for more detailed study of the spatial dependence of the difference. These lines are chosen to go through the rotation centers of the flow rolls. The error bars in \figref{comp2}(b) show the $1\sigma$ error of $\Dup$, taking into account both the SEM for the experimental measurements, \figref{MeasStats}(b), and the error of the analytical prediction (4~\%) inherited from the derived value for $E_\mathrm{ac}^{0.5~\SImum}$. The experimental and analytical velocities do not agree within the error of $\Dup$, moreover, a trend of the experimental velocities being larger than the analytical predictions is seen.

\begin{figure}
\centering
\includegraphics[width=\columnwidth]{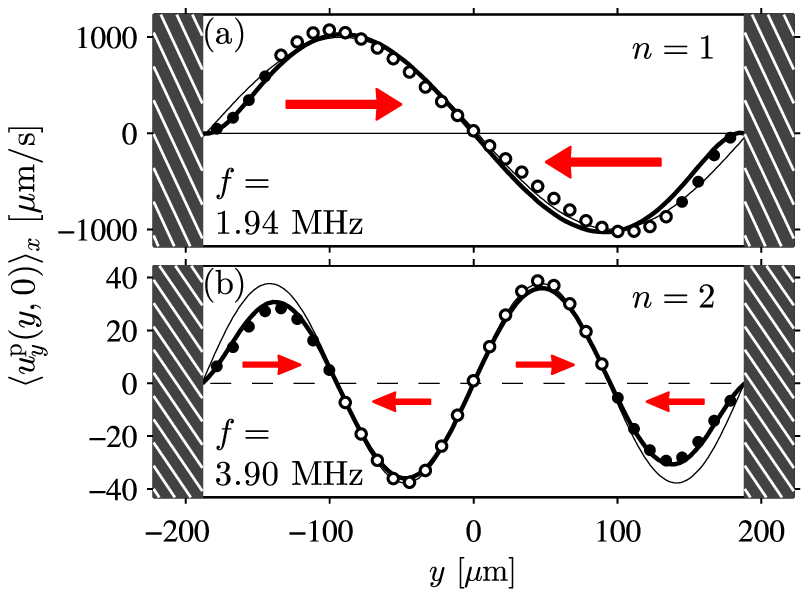}
\caption{\figlab{wall_effects} (Color online) Experimental data from Ref.~\cite{Barnkob2012a} compared with the theoretical predictions of \eqsref{up}{upRect}. $\SImu$PIV measurements, in the center plane $z=0$, of the $y$-component of the acoustophoretic velocity $\avr{\up_y(y,0)}^{{}}_x$ (open and closed dots) for 0.6-$\SImum$-diam polystyrene particles in water, small enough that streaming dominates and $\upvec \approx \avr{\vvvII}$. The observed motion (red arrows) in (a) and (b) resembles the analytical results shown in \figref{streaming_par_plates}(b) and (c), respectively.
For each value of $y$, the measured velocity $\up_y$ is averaged along the $x$-coordinate, with resulting SEM smaller than the size of the dots. The sinusoidal parallel-plate prediction (thin line), \eqref{up}, is fitted to the data points far from the side walls (open dots), while the rectangular-channel prediction (thick line), \eqref{upRect}, is fitted to all data points (open and closed dots). In both fit the acoustic energy, $\Eac$, is treated as a free parameter. (a) The half-wave resonance $\lambda/2 = w$ $(n=1)$ with $f=1.940~\SIMHz$ and $\Upp = 1$~V. (b) The full-wave resonance $\lambda = w$ $(n=2)$ with $f=3.900~\SIMHz$ and $\Upp = 1$~V.
}
\end{figure}

A further comparison between the analytical model presented in this paper and experimental measurements on 0.6-$\SImum$-diam polystyrene particles from Ref.~\cite{Barnkob2012a} is shown in \figref{wall_effects}.
These particles are dominated by the drag from the acoustic streaming, and in this comparison we are only interested in studying how the side walls influence the shape of $\up_y(\yti,0)$, \eqref{upRect}. Consequently, the amplitude of the streaming velocity, and thus the acoustic energy density, is treated as a fitting parameter.
The experimental results support our analytical prediction \eqref{upRect} (thick line) for the rectangular channel with side walls, which shows a suppression of $\up_y$ near the walls compared to the sinusoidal form of $\up_y$ in \eqref{up} (thin line) predicted for the parallel-plate channel without side walls. This is particularly clear for the full wave resonance $\lambda = w$ $(n=2)$ \figref{wall_effects}(b). The difference in the amplitude of $\up$ between \figref{wall_effects}(a) and (b) is due to differences in the resonance modes, $\ie$ \ $\Eac$ is not the same even though $\Upp$ is.

\section{Discussion}
\seclab{discussion}

The comparison shows good agreement between the experimental measurements and the analytical prediction of the streaming-induced particle velocities. The qualitative agreement is seen in \figref{comp1} for the two-dimensional topology of the particle motion, and in \figref{wall_effects} for the effect of the side walls.
Quantitatively, the experimental and analytical results agree within a mean relative difference of approximately $20\:\%$, a low deviation given state-of-the-art in the field. However, as illustrated by the statistical analysis in \figref{comp2}, the differences $\Delta \up$ are larger than the estimated 1$\sigma$-errors. This could indicate a minor systematic error in the experimental procedure or in the theoretical model, or be due to underestimation of the experimental error involved in the analytical prediction.

In the 5-$\SImum$-diam-particle experiment the acoustic energy density is determined using only five particle trajectories close to the channel center $z=0$. This is reasonable as the error of the calculated energy density is relatively low (3\%), however, a calculation based on more particle trajectories would be desirable. This can be realized through more experimental repetitions or through implementation of the 2D-dependence of the wall-enhanced drag force, allowing for use of off-center particle trajectories.
One source of error that has not yet been discussed arises from the assumption made that the acoustic energy density, and thus the acoustic forces, does not depend on the $x$-position in the investigated field of view.
In the same setup, Augustsson \etal\ \cite{Augustsson2011} observed negligible field gradients in the $x$-direction in some field of views and significant ones in others.
This inhomogeneity was considered here when making the measurements: we made sure to check that the five 5-$\SImum$-diam-particle trajectories sample the $x$-range reasonably well and exhibit only negligible variations in the acoustic energy density as a function of $x$-position.

In the 0.5-$\SImum$-diam-particle experiment the statistics and sampling of the $x$-range are good, however they could still be improved to achieve better statistics close to the walls. The relative positions of the 0.5-$\SImum$-diam particles are accurately determined by use of the APTV technique, whereas the absolute position in the channel, which was used to compare with theory, is difficult to determine precisely and might also be improved. Furthermore, accurate measurements of the channel dimensions are also important, as these are key parameters in the theoretical model.

The analytical model could be improved in several ways.
The treatment of the liquid could be extended by including thermal dependence of more material parameters such as the specific heat capacity ratio $\gamma$, thermal expansion $\alphap$, compressibility $\kappas$, and speed of sound $\cO$.
The influence of the surrounding chip material could be included, thereby relaxing the assumptions of infinite acoustic impedance (ideal reflection) and infinite thermal conduction (ideal heat sink) of the channel walls. Solving the full elastic wave problem in the whole chip is beyond analytical solutions, but is, however, possible with numerical models. This might be necessary to achieve accurate quantitative agreement between theoretical predictions and experiments.
Furthermore, the analytical and numerical models assume an ideal rectangular channel cross section, which is crucial since the generating mechanism for the acoustic streaming takes place within the $\SImum$-thin acoustic boundary layer. Even small defects, such as uneven surfaces on the $\SImum$-scale, might lead to changes in the acoustic streaming velocity field.


\section{Conclusions}
\seclab{conclusions}

In this work we have for a rectangular microchannel derived an analytical expression for the acoustophoretic velocity of microparticles resulting from the acoustic radiation force and the acoustic streaming-induced drag force, and we have successfully compared it with a direct numerical solution of the governing equations. We have also accurately measured 3D trajectories of 0.5-$\SImum$-diam and 5-$\SImum$-diam particles in an acoustically actuated microchannel, with an average relative experimental error of $4\:\%$ for the 0.5-$\SImum$-particle velocities. This allowed us to perform a quantitative comparison in 3D between theory and experiments of streaming-induced particle velocities in a rectangular channel. The analytical derivation successfully predicted the measured streaming-induced 0.5-$\SImum$-diam-particle velocities, with qualitative agreement and quantitative differences around $20\:\%$, a low deviation given state-of-the-art in the field.
This shows that the time-averaged second-order perturbation model of the governing equations yields an adequate description of the acoustophoretic particle motion.

The differences between the theoretical prediction and the experimental results emphasize the need for further extensions of the analytical model, along with improved numerical simulations~\cite{Muller2012}.
Aiming for more detailed quantitative studies of acoustophoresis, the results also stress the need for improved accuracy of the measurements of the channel dimensions and the absolute positions of the particles in the microchannel. The trinity of analytical, numerical, and experimental studies of the acoustophoretic particle motion enhance the understanding of acoustophoresis and supports a more elaborate and broader application of acoustophoresis.

\begin{acknowledgments}
This work was supported by the Danish Council for Independent Research, Technology and Production Sciences, Grants No.~274-09-0342 and No.~11-107021, the German Research Foundation (DFG), under the individual grants program KA 1808/12-1, the Swedish Governmental Agency for Innovation Systems, VINNOVA, the program Innovations for Future Health, Cell CARE, Grant No.~2009-00236, and the Swedish Research Council, grant. no. 621-2010-4389.
\end{acknowledgments}



%

\end{document}